# Reversible Logic Based Concurrent Error Detection Methodology For Emerging Nanocircuits

Himanshu Thapliyal and Nagarajan Ranganathan
Department of Computer Science and Engineering,
University of South Florida, Tampa, FL, USA
Email: {hthapliy, ranganat}@cse.usf.edu

*Abstract*—**Reversible logic has promising applications in emerging nanotechnologies, such as quantum computing, quantum dot cellular automata and optical computing, etc. Faults in reversible logic circuits that result in multi-bit error at the outputs are very tough to detect, and thus in literature, researchers have only addressed the problem of online testing of faults that result single-bit error at the outputs based on parity preserving logic. In this work, we propose a methodology for the concurrent error detection in reversible logic circuits to detect faults that can result in multi-bit error at the outputs. The methodology is based on the inverse property of reversible logic and is termed as 'inverse and compare' method. By using the inverse property of reversible logic, all the inputs can be regenerated at the outputs. Thus, by comparing the original inputs with the regenerated inputs, the faults in reversible circuits can be detected. Minimizing the garbage outputs is one of the main goals in reversible logic design and synthesis. We show that the proposed methodology results in 'garbageless' reversible circuits. A design of reversible full adder that can be concurrently tested for multi-bit error at the outputs is illustrated as the application of the proposed scheme. Finally, we showed the application of the proposed scheme of concurrent error detection towards fault detection in quantum dot cellular automata (QCA) emerging nanotechnology.**

## I. INTRODUCTION

Reversible logic is a promising computing paradigm in which there is a one-to-one mapping between the input and the output vectors. *Reversible logic can help in realizing dissipation less computing [19].* The major goal in reversible logic design and synthesis is to minimize the number of garbage outputs and the quantum cost. Garbage outputs refer to the unutilized outputs that are not used as primary outputs and which cannot be used as inputs for new computation, and are only needed to maintain one-to-one mapping. Reversible logic has extensive applications in nanotechnologies, such as quantum computing, quantum dot cellular automata (QCA) and optical computing. Quantum computing is based on reversible gates performing unitary operations, which are reversible in nature [5]. There are existing works on offline testing of reversible logic circuits, the details of which can be found in [3, 9-11]. Recently, concurrent error detection in reversible logic circuits has also attracted the attention of researchers. Concurrent error detection (CED) is defined as the property of circuits in which faults/errors can be detected at run time while the circuit is performing the normal operations. In [1, 2], online testing methodologies for reversible circuits are proposed that can address online testing of faults that results in single bit error at the outputs of the reversible circuits. In [6, 15], the application of conservative reversible logic for concurrent testing of single missing/additional cell defect in quantum dot cellular automata circuits is demonstrated. All these existing works in literature are suitable for detection of faults that result in single bit error at the output of the reversible logic circuits [14]. In this work, we propose a methodology of concurrent error detection in reversible logic circuits based on the inverse property of reversible logic and is called inverse and compare. By using the inverse property of reversible logic, inputs can be regenerated at the outputs. Thus, the regenerated inputs can be compared with the original inputs for concurrent detection of fault (permanent or transient in nature) that result in single bit error or multi-bit error at the outputs of the reversible logic circuits. We showed that the proposed scheme is especially suitable for reversible logic design and synthesis as it produces the *'garbageless'* design. *Minimization of the garbage outputs is one of the major goals in reversible logic design and synthesis.* To illustrate the applications of the proposed inverse and compare scheme of CED, the design of error detectable 1-bit reversible full adder is illustrated as an example circuit. Finally, the application of the proposed concurrent error detection strategy is demonstrated for fault detection in QCA emerging nanotechnology implemented using the reversible logic gates.

## II. PROPOSED SCHEME OF CONCURRENT ERROR DETECTION IN REVERSIBLE LOGIC CIRCUITS

In the existing works on reversible logic, the parity mismatch between the inputs and the outputs is used for concurrent error detection. Hence, the existing works are limited for single bit error at the outputs of the reversible logic circuits.

### A. Inverse and Compare Scheme

For each reversible gate R that maps each input vector X to a unique output vector Y (producing R(X)=Y), there also exist a inverse reversible gate R' which maps each input vector Y to a unique output vector X (producing R'(Y)=X). Thus, the cascading of a reversible gate with its inverse will regenerate the inputs. Figure 1 shows an example of n inputs reversible gate cascaded with its inverse leading to regeneration of the inputs. The cascading of a reversible logic gate with its inverse will minimize the garbage outputs. This is because regeneration of the inputs results in *'garbageless'* reversible circuits as inputs are not considered as garbage signals [4]. We observed that we can use this property for concurrent detection of faults in reversible logic circuits that results in multi-bit errors at the outputs. It is to be noted that fan-out is not allowed in reversible logic. Thus, the fan-out is avoided by using the reversible Feynman gate (FG). Feynman gate is a 2 inputs 2 outputs reversible gate having inputs to outputs mapping as P=A and Q= A⊕B where P and Q are the outputs, and A and B are the inputs, respectively [7]. Feynman gate is shown in Fig. 2.a. Thus in Feynman gate setting the input B to 0 as shown in Fig. 2.b will copy the input A to both the outputs P and Q thus avoids the fan-out problem.

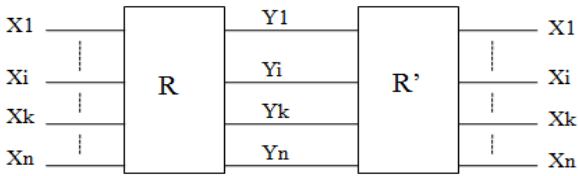

Figure 1. Cascading of a reversible gate R with its inverse R'

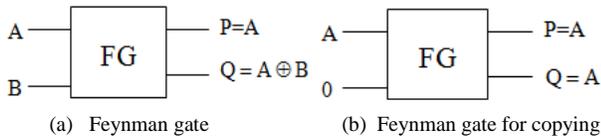

(a) Feynman gate  (b) Feynman gate for copying

Figure 2. Feynman gate and its use for avoiding the fan-out

Figure 3 shows the proposed concurrent error detection methodology. In Fig. 3, the garbage outputs are directly passed to the inverse gate R' to regenerate the inputs. The primary output is represented by Yk where $1 \leq k \leq n$. Hence, it is passed through the Feynman gate (FG) to have the copies of the output Yk, so that one copy can be used as the primary output and the other copy can be passed to the inverse gate (R') to regenerate the input Xk. Thus, in case of faults in either R or its inverse R' or in both of them, there will be a mismatch between the regenerated inputs compared to the original inputs. The regenerated inputs can be compared with the original inputs using a comparator to detect the faults as shown in Fig.3 (*the comparator is assumed fault free as generally considered in redundancy based error detection schemes [12]*). To simplify the discussion, we are assuming that FG gates used for avoiding the fan-out will be fault tolerant in nature. Further, fault can be easily detected in Feynman gate because when input B is set to 0 in Feynman gate, input A and outputs P and Q should have same value. Any mismatch in the values of input A and the outputs P and Q when input B=0, will result in fault detection in the Feynman gate. Almost all nanotechnologies except quantum computing that have applications of reversible logic allows the fan-out, so Feynman gates will not be required in those nanotechnologies to copy the useful outputs. Thus *the proposed scheme provides the 'garbageless' reversible logic circuits with the primary advantage of concurrent error detection.*

B. *Comparison of the proposed scheme of Concurrent Error Detection With Duplication Based Approach*

The duplicate and the compare scheme that is widely used in literature for concurrent error detection has the limitation that the it won't work for the cases in which both the monitored as well as the duplicated circuit have the identical errors [12]. In this case errors can go unnoticed. The inverse and compare scheme will be beneficial in this case as it can also detect the errors when the monitored as well as the inverse circuit have the identical errors. This is because in the inverse and compare scheme we regenerate the inputs so errors in either montiored or inverse circuit or in both of them cannot go unnoticed. Further, since the proposed scheme results in garbageless reversible circuits it is especially suitable for reversible computing as the primary goal in reversible logic design and synthesis is to minimize the number of garbage outputs.

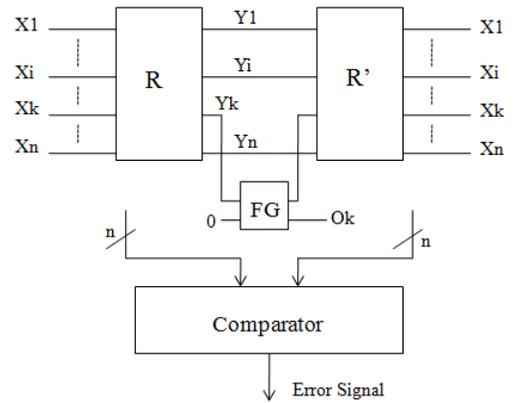

Figure 3. Proposed scheme for concurrent error detection for multi-bit errors at the outputs (fan-out is avoided by using Feynman gate (FG)).

III. EXAMPLE OF CONCURRENT ERROR DETECTION IN REVERSIBLE FULL ADDER

We demonstrate the application of the proposed approach of concurrent error detection to design a concurrently testable reversible full adder in which multi-bit error at the outputs can be detected. In [2], a 4x4 (4 inputs: 4 outputs) reversible gate called OTG gate shown in Fig. 4 is proposed. The truth table of the OTG gate is shown in Table I. OTG gate has the quantum cost of 6 since it is designed from 6 2x2 reversible gates as shown in Fig. 5. The quantum cost of a reversible gate is the number of 1x1 and 2x2 reversible gates or quantum logic gates required in designing it. The quantum cost of all reversible 1x1 and 2x2 gates is taken as unity [6]. Thus any reversible gate is realized by using 1x1 NOT gate, and 2x2 reversible gates such as V and $V^+$ (V is a square-root-of NOT gate and $V^+$ is its hermitian) and Feynman gate which is also known as Controlled NOT gate (CNOT). In simple terms, the quantum cost of a reversible gate can be calculated by counting the number of V, $V^+$ and CNOT gates used in implementing it except in few cases. More details on quantum cost calculation can be found in [18].

OTG gate has the special property that it can work singly as a reversible full adder as shown in Fig. 6. A reversible full adder can be designed using OTG gate with the quantum cost of 6, and 2 garbage outputs. We are demonstrating the application of the proposed approach to design a concurrently testable reversible full adder in which multi-bit errors at the outputs can be detected. This requires a 4x4 reversible gate that can work as the inverse of the OTG gate. We called the inverse of the OTG gate as the IOTG gate. In order to derive the logic equations of the IOTG gate, we performed the reverse mapping of the OTG gate outputs working as inputs to the IOTG gate. This will regenerate the inputs of the OTG gate at the outputs of the IOTG gate leading to the truth table of the IOTG shown in Table II. From the truth table, we derived the logical input and output mapping of the IOTG gate shown in Fig. 7. Figure 8 shows the proposed quantum implementation of the IOTG gate which concludes that the IOTG gate has the quantum cost of 6.

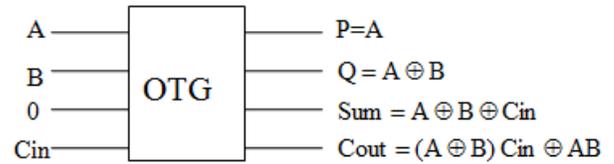

Figure 6. OTG gate as a reversible full adder

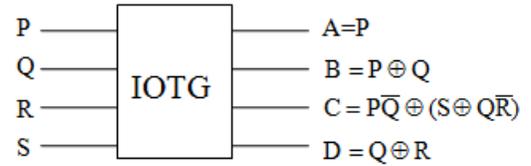

Figure 7. IOTG gate

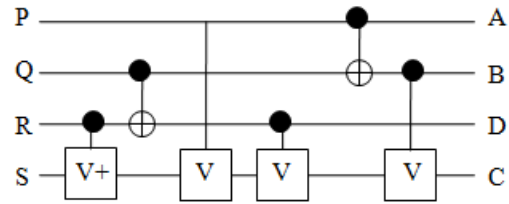

Figure 8. Quantum Implementation of IOTG gate

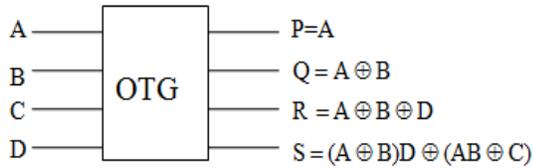

Figure 4. OTG gate

TABLE I. TRUTH TABLE OF OTG GATE

| A | B | C | D | P | Q | R | S |
|---|---|---|---|---|---|---|---|
| 0 | 0 | 0 | 0 | 0 | 0 | 0 | 0 |
| 0 | 0 | 0 | 1 | 0 | 0 | 1 | 0 |
| 0 | 0 | 1 | 0 | 0 | 0 | 0 | 1 |
| 0 | 0 | 1 | 1 | 0 | 0 | 1 | 1 |
| 0 | 1 | 0 | 0 | 0 | 1 | 1 | 0 |
| 0 | 1 | 0 | 1 | 0 | 1 | 0 | 1 |
| 0 | 1 | 1 | 0 | 0 | 1 | 1 | 1 |
| 0 | 1 | 1 | 1 | 0 | 1 | 0 | 0 |
| 1 | 0 | 0 | 0 | 1 | 1 | 1 | 0 |
| 1 | 0 | 0 | 1 | 1 | 1 | 0 | 1 |
| 1 | 0 | 1 | 0 | 1 | 1 | 1 | 1 |
| 1 | 0 | 1 | 1 | 1 | 1 | 0 | 0 |
| 1 | 1 | 0 | 0 | 1 | 0 | 0 | 1 |
| 1 | 1 | 0 | 1 | 1 | 0 | 1 | 1 |
| 1 | 1 | 1 | 0 | 1 | 0 | 0 | 0 |
| 1 | 1 | 1 | 1 | 1 | 0 | 1 | 0 |

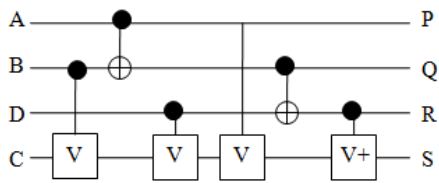

Figure 5. Quantum Implementation of OTG gate

TABLE II. TRUTH TABLE OF IOTG GATE

| P | Q | R | S | A | B | C | D |
|---|---|---|---|---|---|---|---|
| 0 | 0 | 0 | 0 | 0 | 0 | 0 | 0 |
| 0 | 0 | 0 | 1 | 0 | 0 | 1 | 0 |
| 0 | 0 | 1 | 0 | 0 | 0 | 0 | 1 |
| 0 | 0 | 1 | 1 | 0 | 0 | 1 | 1 |
| 0 | 1 | 0 | 0 | 0 | 1 | 1 | 1 |
| 0 | 1 | 0 | 1 | 0 | 1 | 0 | 1 |
| 0 | 1 | 1 | 0 | 0 | 1 | 0 | 0 |
| 0 | 1 | 1 | 1 | 0 | 1 | 1 | 0 |
| 1 | 0 | 0 | 0 | 1 | 1 | 1 | 0 |
| 1 | 0 | 0 | 1 | 1 | 1 | 0 | 0 |
| 1 | 0 | 1 | 0 | 1 | 1 | 1 | 1 |
| 1 | 0 | 1 | 1 | 1 | 1 | 0 | 1 |
| 1 | 1 | 0 | 0 | 1 | 0 | 1 | 1 |
| 1 | 1 | 0 | 1 | 1 | 0 | 0 | 1 |
| 1 | 1 | 1 | 0 | 1 | 0 | 0 | 0 |
| 1 | 1 | 1 | 1 | 1 | 0 | 1 | 0 |

Figure 9 shows the proposed design of 1 bit reversible full adder by combining the OTG gate and the IOTG gate in which multi-bit error at the output can be concurrently detected. In Fig. 9, Feynman gates (FG) are used to avoid the fan-out. It can be seen that the inputs to OTG gate are regenerated at the outputs of IOTG gate. The proposed approach of concurrent error detection is compared with the existing approach in literature [1] that works for single-bit error at the output. The online testing methodology in [1] is based on a combination of

R1 gate along with R2 gate (a 4*4 Feynman Gate). R1 gate is shown in Fig. 10, and to the best of our knowledge the quantum cost of R1 gate is not known. We computed the lower bound on the quantum cost of R1 gate. We can see from the output equations of the R1 gate that there are 7 independent XOR operations and 2 independent AND operations. XOR operation which are counted once and are also used in other places is not counted multiple times, for example, P=A⊕C and R=A⊕B⊕C will have only 2 XOR operations as R can be computed as R=P⊕B. Each XOR operation can be designed by a 2x2 Feynman gate which has the quantum cost of 1, thus R1 gate will have a minimum quantum cost of 7. The minimum quantum cost of testable block formed by R1 gate and R2 gate (4x4 Feynman gate) will be 7+3=10, which is the summation of the quantum cost of R1 gate and 4x4 Feynman gate.

In the delay computation, we have taken the delay of each reversible gate as 1 unit which is fair for comparison purpose as the IOTG and OTG gates are computationally less complex compared to R1 gate. Table III shows the comparison of the proposed design of concurrently testable reversible full adder shown in Fig. 13 with the testable full adder design proposed in [1]. The results shows the significant improvement of 50%, 100%, 62.5% and 33.33% in terms of number of reversible gates, garbage outputs, unit delay and quantum cost, respectively, compared to the design in [1]. Further, our approach has the primary advantage of concurrent detection of multi-bit errors at the outputs.

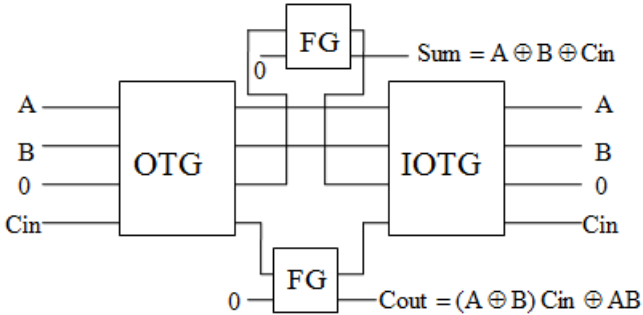

Figure 9. Proposed concurrently testable reversible full adder

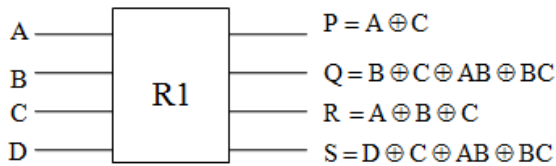

Figure 10. R1 gate proposed in [1] for online testing

TABLE III. A COMPARISON OF TESTABLE REVERSIBLE FULL ADDERS

|  | Gates | Garbage Outputs | Unit Delay | Quantum Cost |
|---|---|---|---|---|
| Full Adder [1] | 8 | 3 | 8 | 30 |
| Proposed Design | 4 | 0 | 3 | 20 |
| *Improvement in %* | 50 | 100 | 62.5 | 33.33 |

## IV. APPLICATION TO EMERGING NANOTECHNOLOGIES

To demonstrate the application of the proposed approach of concurrent error detection in emerging nanotechnologies, we choose quantum dot cellular automata (QCA) nanotechnology as an example since reversible logic has potential applications in QCA computing [6, 15]. Quantum dot cellular automata (QCA) is one of the emerging nanotechnologies in which it is possible to achieve circuit densities and clock frequencies much beyond the limit of existing CMOS technology. QCA has significant advantage in terms of power dissipation as it does not have to dissipate all its signal energy hence considered as one of the promising technologies to achieve the thermodynamic limit of computation. The basic QCA logic devices comprise the majority voter (MV), the inverter (INV), binary wire and the inverter chain [8, 16].

### A. Reversible gates in QCA Computing

QCA computing is based on majority voting, thus recently two new 3x3 (3 inputs: 3 outputs) reversible gates QCA1 and QCA2 suitable for majority based QCA computing are proposed [13]. The reversible QCA1 gate can be described as mapping (A, B, C) to (P=MV(A,B,C), Q=MV(A,B,C'), R=MV(A',B,C), where A, B, C are inputs and P, Q, R are outputs, respectively. The reversible QCA2 gate can be described as mapping (A, B, C) to (P=MV(A,B,C), Q=MV(A,B,C'), R=MV(A',B,C'), where A, B, C are inputs and P, Q, R are outputs, respectively. Figure 11 shows the QCA1 and QCA 2 gates. Since QCA1 and QCA2 are most useful for QCA computing once we have the inverse of QCA1 and QCA2 gates, the proposed method of concurrent error detection can be applied to QCA computing. In this work, we have called the inverse of QCA1 gate as IQCA1, while the inverse of QCA2 gate is called IQCA2.

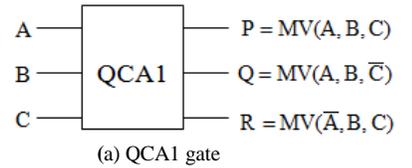

(a) QCA1 gate

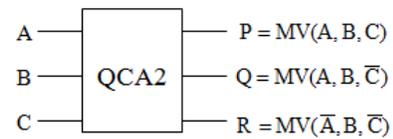

(b) QCA2 gate

Figure 11. QCA1 and QCA 2 reversible gates [13]

### B. Proposed Inverse QCA1 gate (IQCA1) and Inverse QCA2 (IQCA2) gates

In order to derive the inverse of QCA1 gate (IQCA1) we used the truth table of QCA1 gate. The truth table of the QCA1 gate is shown in Table IV. From Table IV, we derive the truth table of the IQCA1 gate as shown in Table V. From Table V, IQCA1 can be described as mapping (P, Q, R) to (A=MV(P,Q,R'), B=MV(P,Q,R), C=MV(P,Q',R), where P, Q,

R are inputs and A, B, C are outputs, respectively. Figure 12 illustrates the IQCA1 gate. Similarly for deriving the inverse of QCA2 gate (IQCA2 ), we used the truth table of QCA2 gate and derived the IQCA2 logic equations as mapping (P, Q, R) to (A=MV(P,Q,R'), B=MV(P,Q,R), C=MV(P,Q',R')), where P, Q, R are input and A, B, C are output, respectively. Figure 13 shows the IQCA2 reversible gate.

Once we have the inverse of QCA1 and QCA2 gates, they can be cascaded with them, respectively, to regenerate the inputs for concurrent error detection of multi-bit error at the outputs in QCA computing. In Fig. 14, an example of the proposed approach is shown for QCA computing by combining QCA1 and IQCA1 together. In QCA computing, fan-out is allowed; hence in Fig.14 we don't require the Feynman gates in the design to avoid the fan-out problem. Thus, implementing a reversible gate in QCA and cascading with its inverse will result in concurrent detection of faults in QCA circuits. The proposed methodology is independent of the reversible gate. Thus any reversible gate along with its inverse implemented in QCA technology can be used for concurrent detection of multiple faults in QCA circuits. *We want to emphasize a very special characteristic of the proposed approach of concurrent testing based on reversible logic for QCA computing. By cascading a reversible logic gate with its inverse will result in dissipation less QCA circuit, as all inputs are regenerated that results in no information loss [17].*

TABLE IV.  TRUTH TABLE OF QCA1

| A | B | C | P | Q | R |
|---|---|---|---|---|---|
| 0 | 0 | 0 | 0 | 0 | 0 |
| 0 | 0 | 1 | 0 | 0 | 1 |
| 0 | 1 | 0 | 0 | 1 | 1 |
| 0 | 1 | 1 | 1 | 0 | 1 |
| 1 | 0 | 0 | 0 | 1 | 0 |
| 1 | 0 | 1 | 1 | 0 | 0 |
| 1 | 1 | 0 | 1 | 1 | 0 |
| 1 | 1 | 1 | 1 | 1 | 1 |

TABLE V.  TRUTH TABLE OF IQCA1

| P | Q | R | A | B | C |
|---|---|---|---|---|---|
| 0 | 0 | 0 | 0 | 0 | 0 |
| 0 | 0 | 1 | 0 | 0 | 1 |
| 0 | 1 | 0 | 1 | 0 | 0 |
| 0 | 1 | 1 | 0 | 1 | 0 |
| 1 | 0 | 0 | 1 | 0 | 1 |
| 1 | 0 | 1 | 0 | 1 | 1 |
| 1 | 1 | 0 | 1 | 1 | 0 |
| 1 | 1 | 1 | 1 | 1 | 1 |

V.  CONCLUSIONS

We have demonstrated a new methodology of concurrent error detection in reversible logic circuits. The proposed strategy is based on the inverse property of reversible logic that helps in the regeneration of the inputs. This results in detection of multi-bit errors at the outputs by comparing the original inputs with the regenerated inputs. The inverse and the compare scheme will be able to detect all types of faults in reversible logic circuits. The proposed methodology of concurrent error detection based on property of reversible logic is generic in nature, and will be applicable to any emerging nanotechnology, such as QCA, nano-CMOS designs, which may be susceptible to single or multiple transient and permanent faults. An application of the proposed approach for concurrent error detection in emerging technologies is illustrated for QCA nanotechnology.

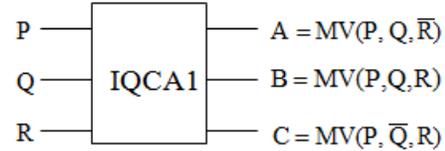

Figure 12. IQCA1  reversible gate

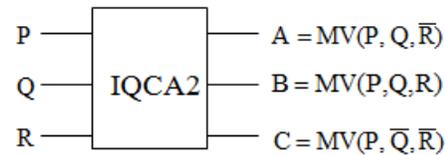

Figure 13. IQCA2 reversible gate

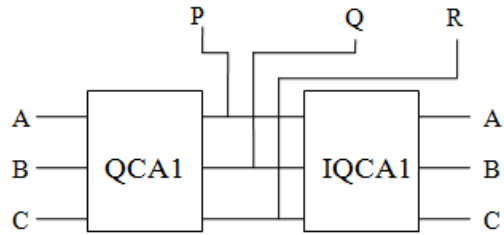

Figure 14. Cascading of  QCA1 and IQCA1  to regenerate the inputs